\documentclass[twocolumn]{aastex631}

\newcommand{\radm}{\rm rad\ m^{-2}}
\newcommand{\kms}{\rm km\ s^{-1}}
\newcommand{\ASKAP}{ASKAP\ J173608.2$-$321635}
\usepackage{gensymb}
\usepackage{multirow}
\usepackage{amsmath}

\begin{document}

\title{Spectrum and polarization of the Galactic center radio transient \ASKAP\ \\ from THOR-GC and VLITE}

\author[0009-0004-9607-721X]{Kierra J. Weatherhead}
\affiliation{Department of Physics and Astronomy, The University of Calgary, 2500 University Drive NW, Calgary AB T2N 1N4, Canada}
\author[0000-0003-2623-2064]{Jeroen M. Stil}
\affiliation{Department of Physics and Astronomy, The University of Calgary, 2500 University Drive NW, Calgary AB T2N 1N4, Canada}
\author[0009-0009-0025-9286]{Michael Rugel}
\affiliation{Center for Astrophysics, Harvard \& Smithsonian, 60 Garden St., Cambridge, MA 02138, USA}
\affiliation{National Radio Astronomy Observatory, 1003 Lopezville Rd, Socorro, NM 87801, USA}
\author[0000-0002-5187-7107]{Wendy M. Peters}
\affiliation{U.S.\ Naval Research Laboratory, 4555 Overlook Ave SW, Washington, DC 20375, USA}
\author[0000-0001-8800-1793]{Loren Anderson}
\affiliation{Department of Physics and Astronomy, West Virginia University, Morgantown, WV 26506, USA}
\affiliation{Adjunct Astronomer at the Green Bank Observatory, P.O. Box 2, Green Bank, WV 24944, USA}
\affiliation{Center for Gravitational Waves and Cosmology, West Virginia University,\\ Chestnut Ridge Research Building, Morgantown, WV 26505, USA}
\author[0000-0003-0410-4504]{Ashley Barnes}
\affiliation{European Southern Observatory (ESO), Karl-Schwarzschild-Stra{\ss}e 2, 85748 Garching, Germany}
\author[0000-0002-1700-090X]{Henrik Beuther}
\affiliation{Max Planck Institute for Astronomy, Königstuhl 17, D-69117 Heidelberg, Germany}
\author[0000-0001-6812-7938]{Tracy E. Clarke}
\affiliation{U.S.\ Naval Research Laboratory, 4555 Overlook Ave SW, Washington, DC 20375, USA}
\author[0000-0001-6010-6200]{Sergio A. Dzib}
\affiliation{Max-Planck-Institut f\"ur Radioastronomie, Auf dem H\"ugel 69, D-53121 Bonn, Germany}
\author[0000-0002-6622-8396]{Paul Goldsmith}
\affiliation{Jet Propulsion Laboratory, California Institute of Technology, 4800 Oak Grove Drive, Pasadena, CA 91109, USA}
\author[0000-0001-6459-0669]{Karl M. Menten}
\affiliation{Max-Planck-Institut f\"ur Radioastronomie, Auf dem H\"ugel 69, D-53121 Bonn, Germany}
\author[0000-0003-1991-370X]{Kristina E. Nyland}
\affiliation{U.S.\ Naval Research Laboratory, 4555 Overlook Ave SW, Washington, DC 20375, USA}
\author[0000-0001-6113-6241]{Mattia C. Sormani}
\affiliation{Department of Physics, University of Surrey, Guildford GU2 7XH, UK}
\affiliation{Universit{\`a} dell'Insubira, via Valleggio 11, 22100 Como, Italy}
\author[0000-0002-1605-8050]{James Urquhart}
\affiliation{Centre for Astrophysics and Planetary Science, University of Kent, Canterbury CT2 7NH, UK}

\begin{abstract}
The radio transient \ASKAP, at the position $(\ell,b)= (356.0872\degree,-0.0390\degree)$, was serendipitously observed by The HI/OH/Recombination Line Survey of the Galactic Center (THOR-GC) at three epochs in March 2020, April 2020 and February 2021. The source was detected only on 2020 April 11 with flux density $20.6 \pm 1.1$ mJy at 1.23 GHz and in-band spectral index $\alpha = -3.1 \pm 0.2$. The commensal VLA Low-band Ionsophere and Transient Experiment (VLITE) simultaneously detected the source at 339 MHz with a flux density $122.6 \pm 20.4\ \rm mJy$, indicating a spectral break below 1 GHz. The rotation measure in April 2020 was $63.9 \pm 0.3\ \radm$, which almost triples the range of the variable rotation measure observed by \citet{wang2021} to $\sim 130\ \radm$. The polarization angle, corrected for Faraday rotation, was $97\degr \pm 6\degr$. The 1.23 GHz linear polarization was $76.7 \pm 3.9\%$ with wavelength-dependent depolarization indicating Faraday depth dispersion $\sigma_\phi = 4.8^{+0.5}_{-0.7}\ \radm$.  We find an upper limit to circular polarization $|V|/I < 10.1\%$. Interpretation of the data in terms of diffractive scattering of radio waves by a plasma near the source indicates electron density and line-of-sight magnetic field strength within a factor 3 of $n_e \sim 2\ \rm cm^{-3}$ and $B_\| \sim 2\times 10^5\ \rm \mu G$. Combined with causality limits to the size of the source, these parameters are consistent with the low-frequency spectral break resulting from synchrotron self-absorption, not free-free absorption. A possible interpretation of the source is a highly supersonic neutron star interacting with a changing environment.
\end{abstract}


\section{Introduction} \label{sec:intro}

Increases in survey speed, wide-field capability and capabilities for commensal surveys have increased the chance of detection of transient radio continuum sources with a variety of origins such as Fast Radio Bursts \citep[FRBs;][]{2022A&ARv..30....2P}, Rotating Radio Transients \citep[RRATS;][]{2006Natur.439..817M}, radio afterglows of gamma ray bursts \citep{1997Natur.389..261F}, tidal disruption events \citep{2020ApJ...903..116A}, stellar flares \citep{2010ApJ...712L...5R}, and intermittent pulsars \citep{2006Sci...312..549K}. Establishing basic observational parameters is important to uncover the underlying physical phenomena, but it can be challenging when sources change in an unpredictable way on short time scales. Some radio interferometers are capable of simultaneous wide-field, high time resolution, high angular resolution observations. Examples are the Australian Square Kilometre Array Pathfinder \citep[ASKAP;][]{hotan2021}, the South African Square Kilometre Array precursor MeerKAT \citep{2016mks..confE...1J}, the Low-Frequency Array \citep[LOFAR;][]{vanhaarlem2013}, the Murchison Widefield Array \citep[MWA;][]{tingay2013}, the Allen Telescope Array \citep[ATA;][]{2010ApJ...719...45C}, and the Canadian HI Intensity Mapping Experiment \cite[CHIME;][]{chime2022} with its outrigger stations. Sometimes, surveys that are not designed to observe transients make significant serendipitous observations of transient sources, as in the case of this paper.

The unresolved, erratic radio transient \ASKAP\ was first discovered in January 2020 when data from the Variables and Slow Transients Phase 1 Pilot Survey \citep[VAST-P1;][]{2013PASA...30....6M} were searched for transient sources. This source had not been detected in ASKAP observations prior to this, between April 2019 and October 2019. ASKAP measurements between January 2020 and August 2020 found J173608.2$-$321635 to be variable and also highly circularly polarized. Later measurements by MeerKAT in February 2021 also found the source to be highly linearly polarized. The nature of \ASKAP\ is unclear. \ASKAP\ has similar parameters to other transient sources near the Galactic center, including its steep spectral index that varies from $\alpha=-2.7$ to $\alpha=-5.6$, where $S_\nu\sim\nu^\alpha$ \citep{wang2021}. Some Galactic Center Radio Transients (GCRTs) have also been reported with very steep spectral indices. GCRT J1742$-$3001 was found to have a spectral index $\lesssim-2$ between 235 MHz and 610 MHz by \citet{Hyman_2009}. GCRT J1745$-$3009 has a variable in-band spectral index at 325 MHz ranging from $-4$ to $-13.5$ \citep{Hyman_2007,Roy_2010}.

\citet{wang2021} and \citet{wang2022} found \ASKAP\ to have a high degree of linear polarization, nearly 100\% at 1.6 GHz, and variable circular polarization up to $\sim40\%$ at 0.9 GHz in two separate observations in 2020 and 2021. \citet{Roy_2010} reported variable circular polarization up to 100\% for the GCRT J1745$-$3009.

The outburst of \ASKAP\ that led to its discovery by ASKAP was serendipitously observed with the Karl G. Jansky Very Large Array (VLA) during observations for the extension towards the Galactic center of
The HI/OH/Recombination line survey of the inner Milky Way \citep[THOR;][]{beuther2016}. 

The THOR Galactic Center extension (THOR-GC) covers $-6\degree<l<15\degree$ with $|b|<1.25\degree$. 
Fortuitously, the VLA Low-Band Ionosphere and Transient Experiment (VLITE; \citealt{Clarke+2016})\footnote{\url{https://vlite.nrao.edu}} system was operational at the time, providing us with simultaneous low-frequency data at 339 MHz. In this work we present the new observations of \ASKAP\ from the THOR-GC survey and discuss some possible interpretation in view of the new results.

\begin{deluxetable*}{cccccccccccc}
\tablecolumns{9}
\tablewidth{0pc} 
\tablecaption{ THOR and VLITE flux density and polarization measurements }
\tablehead{Civil Date & Epoch & $\nu$ & $S_\nu$ &  $\alpha$ &  $\Pi_\nu$ & $\theta$ & $RM$ & $\sigma_\phi$ & $V_\nu/I_\nu$\\     
  & (MJD)   & (MHz) &  (mJy)       &             &  (\%) & $(\degree)$ & $(\radm)$ & $(\radm)$ &  (\%) }
\startdata
     \multicolumn{1}{c}{\multirow{2}{*}{2020 March 17}}& \multirow{2}{*}{58925$^{a}$} & 339  & $<34.5$    &  $\ldots$    &  $\ldots$   & $\ldots$  & $\ldots$ &  $\ldots$ &  $\ldots$  \\   
       &  & 1243$^{c,d}$  & $<6.8$    &  $\ldots$    &  $\ldots$   & $\ldots$  & $\ldots$ &  $\ldots$ &  $\ldots$ \\ \cline{1-2}
     \multicolumn{1}{c}{\multirow{2}{*}{2020 April 11}} & \multirow{2}{*}{58950$^{a}$} &  339 & $122.6 \pm 20.4$    &  $-1.0 \pm 0.2$$^{f}$    &  $\ldots$   & $\ldots$ &   $\ldots$ &  $\ldots$ &  $\ldots$  \\   
      &  & 1233$^{c,d}$  & $20.6 \pm 1.1$    &  $-3.1 \pm 0.2$\phantom{$^d$}    &  $76.7 \pm 3.9$   & $-46.7 \pm 0.3$ &   $63.9 \pm 0.3$ &  $4.8^{+0.5}_{-0.7}$ &  $<10.1^{g}$  \\ \cline{1-2}  
     2021 April 28 & 59332$^{b}$ & 1435$^{c,e}$  & $<11.4$    &  $\ldots$    &  $\ldots$   & $\ldots$   & $\ldots$ &  $\ldots$ &  $\ldots$  \\   
\enddata 
\tablenotetext{a}{C-configuration}
\tablenotetext{b}{D-configuration}
\tablenotetext{c}{Frequencies from 2020 March 17 and 2021 April 28 are derived from multi-frequency synthesis and frequency from 2020 April 11 is derived from $RM$ synthesis}
\tablenotetext{d}{Includes measurements from subbands centered at 1.051 and 1.435 GHz}
\tablenotetext{e}{Includes measurements from subbands centered at 1.435 GHz}
\tablenotetext{f}{Spectral index between 339 MHz and 1 GHz}
\tablenotetext{g}{Upper limit for circular polarization at 1.248 GHz}

\label{results-tab}
\end{deluxetable*}

\section{Observations and data reduction} \label{sec:observatons}
\subsection{THOR-GC}

The HI/OH/Recombination line Survey  of the inner Milky Way \citep[THOR][]{beuther2016} is an L-band (1-2 GHz) survey with the Jansky Very Large Array (VLA) of the interstellar medium (ISM) in the inner Galaxy, with separate data products for spectral lines of the HI 21 cm line, four 18 cm OH lines, several hydrogen recombination lines and the continuum. Details of the observational setup, calibration and imaging were described by \citet{beuther2016} and in particular for the continuum polarization by \citet{shanahan2022}. THOR-GC (VLA project 20A-160) is an extension of the THOR survey towards and across the Galactic center region, with the same data products. A notable difference between THOR and THOR-GC is that THOR was observed in C-configuration only, using D-configuration data from the VLA Galactic Plane Survey \citep[VGPS,][HI line and 1.4 GHz continuum only]{stil2006} that were observed before the upgrade of the VLA. THOR-GC observed \ASKAP\ in C-configuration on 2020 March 17 and 2020 April 11 and with a short snapshot in D-configuration on 2021 April 28.  

Calibration and imaging were performed in the CASA environment 6.5.0 \citep{casa2022} following standard procedures. THOR-GC includes 6 of the continuum subbands used in THOR. In this work, we present observations from subbands centered at 1.05 and 1.44 GHz, each with a bandwidth of 128 MHz. The spectrum of the source was so steep that it was not detected in our higher frequency subbands.

The Stokes $I$ visibility data were averaged to 20 MHz channels, while the Stokes $Q$ and $U$ visibility data were averaged to 4 MHz channels before imaging to improve the signal to noise ratio per channel for cleaning. The averaging reduces the maximum observable Faraday rotation to $2.4 \times 10^3\ \radm$ at 1 GHz, which is sufficient for the purpose of this paper. The Stokes $V$ visibility data were averaged over the two subbands in which \ASKAP\ was visible in Stokes $I$. The $Q$, $U$, and $V$ images were convolved to a common angular resolution of $45\arcsec\times20\arcsec$ using the CASA task imsmooth, with pixels of size $2.5\arcsec\times2.5\arcsec$. 
For each polarization image, a Stokes I image with the same angular resolution was made.

For the analysis of \ASKAP, the observations of the three epochs were imaged separately as outlined in Table~\ref{results-tab}. The final images used in the analysis were made using the surrounding 25 pointings. Each pointing in the mosaic was imaged to a radius where the sensitivity dropped to 20\%. As such, 6 to 11 of the 25 imaged pointings, depending on frequency, covered the location of \ASKAP. The source was near the center of one pointing, at a distance of $3\farcm25$, and the center of the furthest pointing which covered the source at all frequencies used in the analysis was at a distance of $17\farcm96$. The six pointings were each observed on three scans of the sky over a span of 3.59 hours. The average integration time of each scan for the 6 pointings was 1.72 minutes, and each pointing had an average total integration time of 5.17 minutes. We tried to image individual fields to investigate variability on time scales of minutes, but this proved inconclusive because the integration time of the individual snap shots was short and the steep spectrum of the source significantly reduced the effective bandwidth. 

Stokes $Q$ and $U$ spectra were extracted by summation over a box of size $4 \times 4$ pixels centered on \ASKAP. These $Q$ and $U$ spectra were subsequently analyzed with methods designed for broad-band radio polarimetry called Faraday Rotation Measure Synthesis \citep[$RM$ synthesis,][]{brentjens2005} and $QU$ fitting \citep{law2011,osullivan2012}. For a linearly polarized wave with wavelength $\lambda$, traveling through a magnetized plasma with electron density $n_e$ and magnetic field with line-of-sight component $B_\|$, the polarization angle $\theta$ changes by an amount $\phi \lambda^2$ with the Faraday depth $\phi$ defined as
\begin{equation}
\phi = {e^3 \over 2 \pi m_e^2 c^4} \int n_e B_\| \rm{d}l
\label{RM-eq}
\end{equation}
with $e$ the elementary charge, $m_e$ the mass of an electron, and $c$ the speed of light. The integral is performed from the source to the observer, with positive $\phi$ indicating a mean magnetic field directed toward the observer. 

If we encounter a situation in which waves experiencing different amounts of Faraday rotation are blended, the superposition of the polarization states of these waves leads to changes in the observed fractional polarization with wavelength and also deviations from the $\lambda^2$ dependence of the polarization angle that can be observed in polarimetry data with good spectral resolution over a wide wavelength range.

In $RM$ synthesis\footnote{This work uses $RM$-tools \citep{2020ascl.soft05003P} for analysis of polarization data.} \citep{brentjens2005}, the Stokes $Q$ and $U$ spectra are divided by the Stokes $I$ spectrum and combined into the complex polarization $\mathcal{P} = q + i u$, with $q=Q/I$ and $u = U/I$. The division by Stokes $I$ eliminates the power law spectral dependence of optically thin synchrotron emission, which has constant fractional polarization. Introducing a weight function $W(\lambda^2)$ that is zero for any wavelength $\lambda$ for which no measurements are available, the Fourier transform of $P(\lambda^2)W(\lambda^2)$ is the observed Faraday dispersion function $\tilde{\mathcal{F}}$,
\begin{equation}
\tilde{\mathcal{F}} = {\int W(\xi) \mathcal{P}(\xi)\, e^{-2 i \phi \xi}\rm{d} \xi \over \int W(\xi) \rm{d} \xi},
\end{equation}
where $\xi = \lambda^2$ if $\xi > 0$. As no measurements can be made for $\xi \leq 0$, $W(\xi) = 0$ for these values to extend the bounds of the Fourier transform over the required range \citep{brentjens2005}. In this work we apply uniform weights for all wavelengths where measurements were made.

In principle, the Faraday dispersion function is the distribution of polarized intensity as a function of Faraday depth. In practice, $\tilde{F}$ is the convolution of the true Faraday dispersion function with the Fourier transform of $W(\xi)$ and a deconvolution is necessary \citep{heald2009}.

$QU$ fitting achieves the same goals as $RM$ synthesis in $\lambda^2$ space by fitting a model to the complex polarization. A useful model for analysis of the current data is \citep[e.g.][]{burn1966,osullivan2012,2022Sci...375.1266F}
\begin{equation}
\mathcal{P}= P_0 e^{-2 \sigma_\phi^2 \lambda^4} e^{2i(\theta_0 + \phi_0 \lambda^2)}.
\label{depol-eq}
\end{equation}
In this function, the fitting parameters are $P_0$, the (real-valued) fractional polarization in the absence of any Faraday rotation, $\sigma_\phi$, the Faraday depth dispersion, and $\phi_0$, the mean Faraday depth, also referred to as the rotation measure ($RM$). This model is useful when fitting modest depolarization over the observed frequency range, yielding a mean and standard deviation of the Faraday dispersion function. It has been shown to fit the depolarization of FRBs well \citep{2022Sci...375.1266F} although its physical interpretation of a turbulent foreground screen with many unresolved independent cells covering a source \citep{burn1966} does not align well with the compact nature of these sources.  

The standard THOR-GC data products combine the results of both C- and D-configuration observations. In this paper, we investigate a continuum transient source that requires imaging of the individual observing epochs. After some experimentation, it was decided that removing baselines shorter than 500 m (14\% of the visibilities) did not significantly improve the quality of the images, and data obtained with all baselines could be used in the imaging of the L-band data. We note that \ASKAP\ is too faint to detect any absorption in the HI 21-cm line data.

\begin{figure*}
    \centering
    \includegraphics[width=\linewidth]{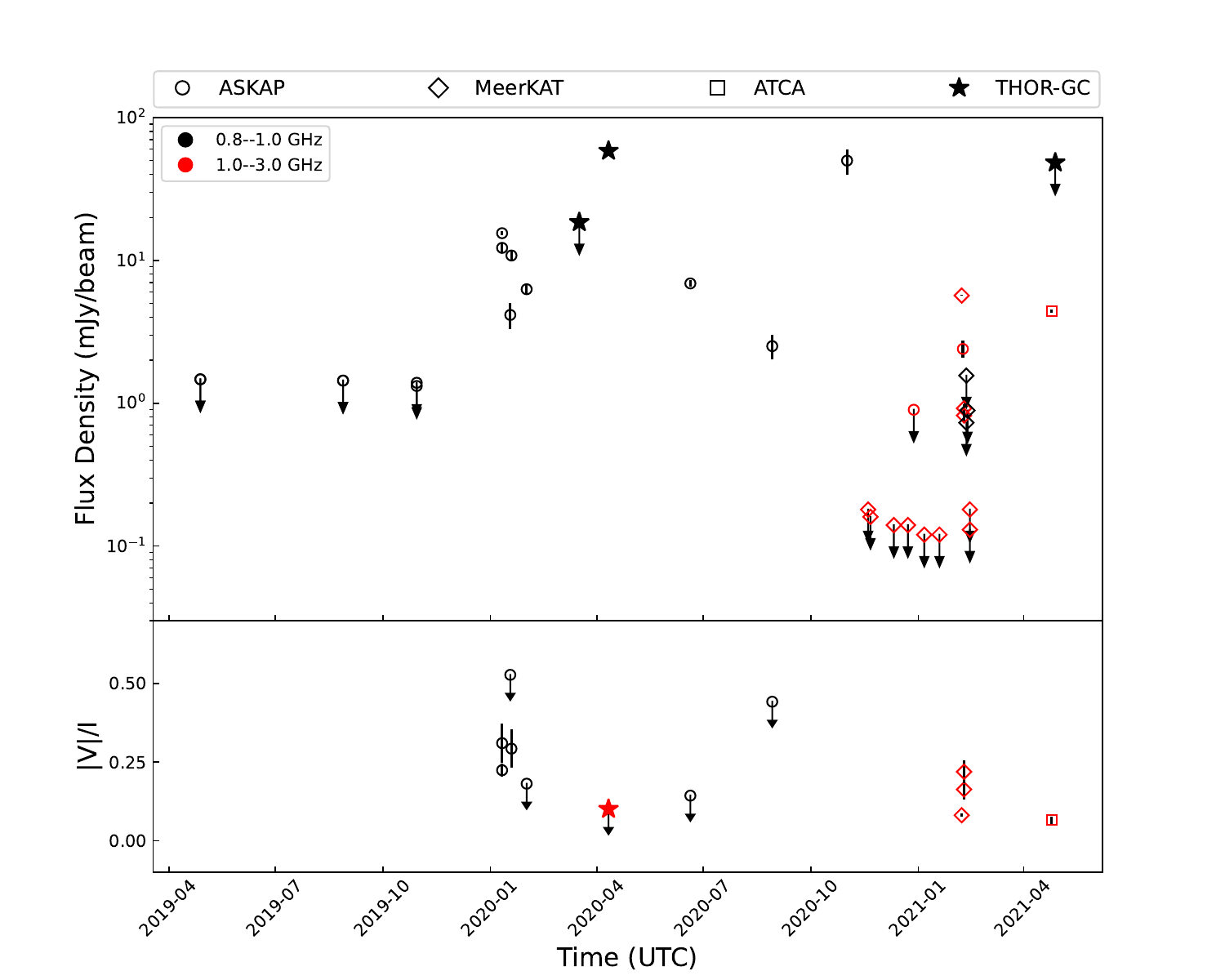}
    \caption{Flux density of \ASKAP\ as a function of time \citep[adapted from][]{wang2021}. The ASKAP, MeerKAT and ATCA flux densities are from \citet{wang2021}. The colour of the symbols indicates the same broad frequency ranges as in \citet{wang2021}. Arrows indicate 3$\sigma$ upper limits. THOR-GC total flux densities and upper limits have been extrapolated to 900 MHz assuming a spectral index of $\alpha=-3.1$. The THOR-GC circular polarization upper limits is for $\nu = 1.23\ \rm GHz$ (Table~\ref{results-tab}). The bottom panel shows fractional circular polarization of the source.}
    \label{fig:fluxvstime}
\end{figure*}

\begin{figure*}
    \centering
    \includegraphics[width=\linewidth]{ASKAP_THOR_VLITE_pos_updated_April8.pdf}
    \caption{Images of \ASKAP\ and surroundings at 1029 MHz from THOR-GC (left) and at 339 MHz from VLITE (right). The red and blue circles are centered on the best-fit positions of \ASKAP\ from THOR-GC and VLITE, respectively. Green circles mark the locations and approximate angular size of supernova remnants in the catalog of \citet{2019JApA...40...36G} updated in 2022 and made available at http://www.mrao.cam.ac.uk/surveys/snrs/. Toward the left border is G356.3$-$0.3 \citep{1994MNRAS.270..847G,2002MNRAS.329..775R}, which is very diffuse and undetected in these narrow-band snap shots. Toward the bottom is G355.6$-$0.0 \citep{1994MNRAS.270..847G,2006JPhCS..54..152R}.  The orange ellipses in the bottom left corners indicate the beam size in THOR (left) and VLITE (right).}
    \label{fig:ASKAP_pos}
\end{figure*}

\begin{figure}
    \centering
    \includegraphics[width=\columnwidth]{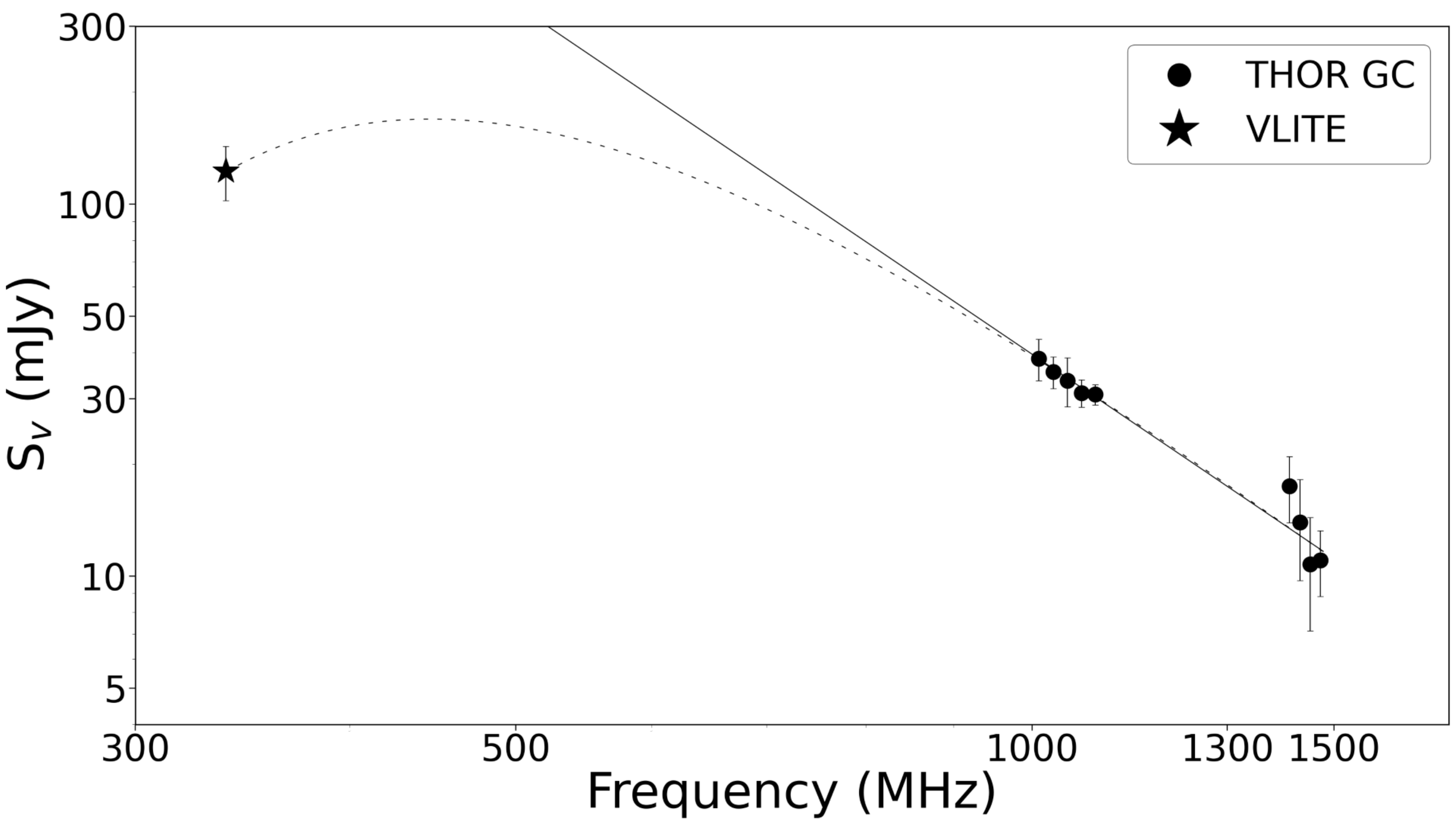}
    \caption{Stokes $I$ flux density of \ASKAP\ from THOR-GC and VLITE observations on 2020 April 11. A power law relation (S$_\nu\sim\nu^\alpha$) is fitted (solid) to the THOR-GC data with $\alpha=-3.1\pm0.2$. A power law relation with free-free absorption is fitted (dashed) to the combined THOR-GC and VLITE data with $\alpha=-3.6\pm0.3$ and $\tau=0.4\pm0.06$ at 1 GHz.}
    \label{fig:specind}
\end{figure}

\subsection{VLITE data processing}
Data from the VLA Low-band Ionosphere and Transient Experiment were recorded simultaneously using 18 antennae during the THOR-GC observations described above in the VLA's C-configuration on 2020 March 17 and 2020 April 11.  The VLITE 2021 D-configuration data did not have sufficient angular resolution to reliably separate the transient source from surrounding structures, and those data are not included here. All VLITE data are processed within a few days of observation by a dedicated calibration and imaging pipeline which combines Python with standard processing tasks found in AIPS \citep{aips} and Obit \citep{obit}.   Full details of the VLITE calibration pipeline are described in \citet{Polisensky+2016}.  For both of the 2020 observing sessions, 3C286 was used for primary calibration, and an NVSS image \citep{Condon1998} was used as a sky model to correct for ionospheric phase contribution in the target direction. The data have a final bandwidth of 34 MHz centered at 339 MHz on both days, and an angular resolution of 86$\arcsec \times 22\arcsec$ at $0.63\degr$ and $60\arcsec \times 26\arcsec$ at $16.9\degr$ on 17 March and 11 April respectively.

In order to match the higher frequency analysis, we used the same 25 surrounding THOR-GC pointings within a radius of 0.72$\degr$ of the target, observed over a span of 3.95 hours. The uv-data for each of these were shifted to a common reference center at the position of the source of interest using the AIPS task `UVFIX', and then combined for each of the two dates. The combined data were manually flagged to remove any remaining radio frequency interference (RFI), and then imaged using the Obit task `MFImage', with a single phase-only self-calibration loop, using a 6-second solution step, to ensure all the data were well-aligned. In order to minimize the contribution of large-scale Galactic structures, during the imaging step we removed the shortest baselines ($< 0.4$ k$\lambda$), and used a slightly uniform weighting scheme (robust factor $-1.5$). 
The images were corrected using VLITE-specific beam models that were averaged to properly account for the original telescope positions.

The final images have rms noise levels at the position of the source of $\sigma= 11.5$, and $10.2$ mJy beam$^{-1}$ on 17 March and 11 April respectively.  The source is not seen on the first day, but is detected at a signal-to-noise ratio of 12 with a flux of $122.6 \pm 20.4\ \rm mJy$ on 11 April.  
\newpage

\section{Results} \label{sec:results}
\subsection{Variability}

\ASKAP\ was not detected in two epochs of THOR-GC on 2020 March 17 and 2021 April 28. Averaging over a single spectral window, 3$\sigma$ upper limits of 13.5 mJy and 4.3 mJy were found for the epoch on 2020 March 17 at 1.051 and 1.435 GHz, respectively, and 11.4 mJy for the epoch on 2021 April 28 at 1.435 GHz. Averaging over two spectral windows for the 2020 March 17 observation, a 3$\sigma$ upper limit of 6.8 mJy was found at 1.243 GHz. The source was strongly detected in our observation on 2020 April 11 with a flux density of $20.6\pm1.1$ mJy at a centroid frequency of 1.233 GHz. Figure \ref{fig:fluxvstime} shows the flux density of the \ASKAP\ detection and upper limits in THOR-GC with measurements made by \citet{wang2021}.

Based on this detection, the best-fit position of the source is RA(J2000) $17^{\text{h}}36^{\text{m}}8 \fs 25 \pm 0 \fs 09$, Dec(J2000) $-32\degr16\arcmin31\farcs71 \pm 1\farcs9$, with Galactic coordinates $(l,b)=(356.0872\degr,-0.0390\degr)$. The best-fit position from the VLITE measurement on the same day is RA(J2000) $17^{\text{h}}36^{\text{m}}8 \fs 21 \pm 0\fs15$, Dec(J2000) $-32\degr16\arcmin41\farcs18\pm8\farcs56$, $(l,b)=(356.0849\degr,-0.0403\degr)$. Figure \ref{fig:ASKAP_pos} shows the surroundings of \ASKAP\ in THOR-GC and VLITE. Flux densities and other source parameters are listed in Table~\ref{results-tab}.

\subsection{X-Ray Observations}
The extended ROentgen Survey with an Imaging Telescope Array \citep[eROSITA;][]{Predehl_2021} all-sky survey includes X-ray observations at soft ($0.2-0.6$ keV), medium ($0.6-2.3$ keV), and hard ($2.3-5.0$ keV) energy bands. The hard energy band is the least susceptible to absorption, and so we use this band in our analysis. eROSITA (data release 1, 31 January 2024) observed this region of the sky in the hard energy band between 2020 March 24 and 2020 March 28 for 77 seconds. There were no counts within 4$\arcsec$ of \ASKAP. This yields an upper limit on the flux of $1.00\times10^{-12}$ erg s$^{-1}$ cm$^{-2}$ from $2.3-5.0$ keV \citep{tubínarenas2024erosita, Merloni_2024}. This corresponds to an upper limit of $1.20\times10^{34}(d/10 \text{ kpc})^2$ erg s$^{-1}$ for the X-ray luminosity at a distance $d$. This upper limit excludes a powerful burst of a magnetar near the Galactic centre \citep[see discussion in ][]{wang2021} at the time of the observation, but the sensitivity of the eROSITA sky survey is much lower than that of the targeted X-ray observations presented by \citet{wang2021}.

\subsection{Spectrum}

The integrated Stokes $I$ spectrum of \ASKAP\ for the 2020 April 11 observation was found to have an in-band spectral index of $\alpha=-3.1\pm0.2$, where $S_\nu\sim\nu^\alpha$. This spectral index is within the range observed by \citet{wang2021}, which decreased from $\alpha=-2.7\pm0.1$ on 2021 February 7 to $\alpha=-5.6\pm0.3$ on 2021 April 25. We used our in-band spectral index to extrapolate L-band flux densities and upper limits to 900 MHz for direct comparison with the results of \cite{wang2021} (Figure~\ref{fig:fluxvstime}). Our detection shows that \ASKAP\ reached a peak brightness at least twice during its 2020/2021 period of activity, and its flux density dropped by a factor $\sim 30$ in between peaks.

The simultaneously measured VLITE flux density at 339 MHz is much lower than the extrapolated L-band spectrum would predict, as shown in Figure \ref{fig:specind}. This indicates a break in the radio spectrum that could be a feature of the mechanism that produces the radio emission. We will discuss synchrotron self-absorption in Section~\ref{sec:discuss_depol}. This break implies that \ASKAP\ had an inverted low-frequency spectrum in contrast to the very steep spectrum of known GCRTs below $\sim500$ MHz \citep{Hyman_2007,Hyman_2009,Roy_2010}.

For our discussion it will be helpful to know what the opacity would be if the break were the result of free-free absorption. A power law relation with free-free absorption ($S_\nu\sim\nu_{\rm GHz}^\alpha e^{-\tau\nu_{\rm GHz}^{-2}}$, where $\tau$ is the opacity at 1 GHz), with $\nu_{\rm GHz} = \nu/1\ \rm GHz$, fits the combined THOR-GC and VLITE data with $\alpha=-3.6\pm 0.3$ and $\tau=0.4\pm 0.06$ at 1 GHz. Assuming a temperature of 10$^4$ K, this free-free opacity corresponds to an emission measure of $\sim1.1\times10^6\ \rm cm^{-6}\ pc$. This is comparable to the emission measure of the Orion Nebula.

\citet{rubin1968} models the total flux density of optically thin free-free emission from regions of varying sizes as
\begin{equation}
    S_\nu=\frac{8.61\times10^{-76}}{D^2\nu^{0.1}} \int n_e n_i T^{-0.35}dV,
\end{equation}
where $D$ is the distance in pc, $\nu$ is the frequency in GHz, $n_e$ and $n_i$ are the electron and ion densities in cm$^{-3}$, $dV$ is a volume element in cm$^{3}$, and $S_\nu$ is in ergs cm$^{-2}$ s$^{-1}$ Hz$^{-1}$. For an optically thin sphere with a 10 pc radius at a distance of 8.3 kpc \citep[the distance to the Galactic center,][]{2021A&A...647A..59G}, assuming a uniform density of 2.0 cm$^{-3}$, which is more normal for the inner Galaxy \citep{2017ApJ...835...29Y}, and a temperature of 10$^4$ K, we find $20$ mJy for the flux density of free-free emission. This flux is high enough that it would have been detected in our measurements, but we do not see any persistent emission at the location of the source. If the radius of the sphere were 1 pc, the expected free-free emission would drop to $2\times10^{-2}$ mJy, which is below our detection threshold. As such, the observed spectral break could be due to a small region of plasma with a density predicted by the modelling, but only if its filling factor in the synthesized beam is very small. We will discuss the possibility that the spectral break is due to free-free absorption further in Section~\ref{discussion-sec}.

\begin{figure*}
    \centering
    \includegraphics[width=\linewidth]{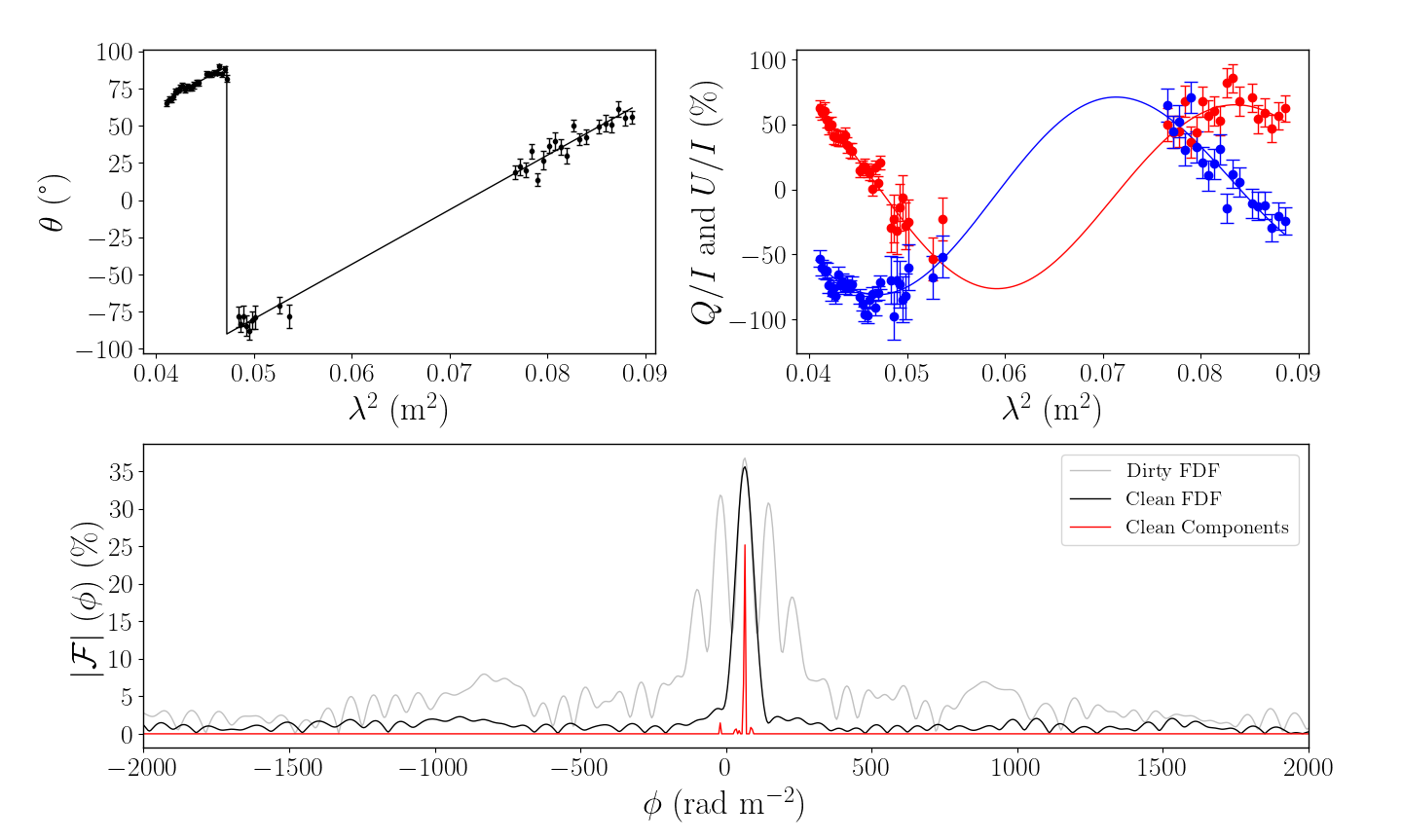}
    \caption{Results of $QU$ fitting and $RM$ clean for \ASKAP\ for the detection in THOR-GC on 2020 April 11. The top left panel shows polarization angle as a function of $\lambda^2$ with the model from $QU$ fitting. The top right panel shows Stokes $Q$ (blue) and $U$ (red) as a function of $\lambda^2$ with the result of $QU$ fitting indicated by the solid lines. The bottom panel shows the dirty Faraday dispersion function (FDF) in gray, clean FDF and $RM$ in black and clean components from $RM$ clean in red.}
    \label{fig:RMsynth QUfitting}
\end{figure*}

\section{Polarization}
\label{pol-sec}
The source was highly linearly polarized ($|L|/I\ \sim\ 76.7\pm 3.9\%$, where $|L|=\sqrt{Q^2+U^2}$) at 1.233 GHz, which is consistent with MeerKAT measurements of $|L|/I \sim 80\%$ \citep{wang2021}. The rotation measure of the source was determined using two  methods: $RM$ synthesis and $QU$ fitting (Figure~\ref{fig:RMsynth QUfitting}). $RM$ synthesis found a rotation measure of $+63.9\pm0.3$ rad m$^{-2}$. The $RM$ value determined by $QU$ fitting was consistent within the margin of error. $QU$ fitting of Equation~\ref{depol-eq} also found the source to be depolarizing, with $\sigma_{\phi}=4.8^{+0.5}_{-0.7}$ rad m$^{-2}$. \citet{wang2021} found a similar depolarization, $\sigma_\phi=5.7$ rad m$^{-2}$, on 2021 February 7, but found the $RM$ to vary from $-11.8\pm0.8$ rad m$^{-2}$ on 2021 February 7 to $-64.0\pm1.5$ rad m$^{-2}$ on 2021 February 9. The THOR-GC measurement increases the $RM$ variability range from $\sim$50 rad m$^{-2}$ to $\sim$130 rad m$^{-2}$.

The polarization angle at the reference frequency 1.23 GHz is $-46\fdg7\pm0\fdg3$. Correcting for Faraday rotation, we find an intrinsic polarization angle $97\fdg1\pm1\fdg9$ from $RM$ synthesis and $96\fdg8_{-3.1}^{+2.7}$ from $QU$ fitting. The uncertainty in the calibration of the absolute polarization angle is approximately $5\degr$ related to hour-angle and baseline dependent effects in the $RL$ phase that are not well understood 
\citep[EVLA memo 205\footnote{https://library.nrao.edu/public/memos/evla/EVLAM\_205.pdf}, see also][]{lacy2020}. The extrapolation to $\lambda = 0$ assumes a linear dependence between polarization angle and $\lambda^2$ with slope $RM$ at all wavelengths. We have no indication to the contrary, but the extrapolated polarization angle is only as good as this assumption \citep{farnsworth2011}. Adding formal and absolute calibration errors in quadrature, we find an intrinsic polarization angle of $97\degr \pm 6\degr$.

The intrinsic polarization angle is related to the projection of the mean magnetic axis of the source on the plane of the sky. \citet{wang2021} could not absolutely calibrate their polarization angles. They reported a reference source that could be used to calibrate the angle a posteriori. Unfortunately, this reference source is well below the detection limit of THOR-GC.

\ASKAP\ was not detected in circular polarization in our observation. We derive a 3$\sigma$ upper limit of 2.0 mJy in Stokes $V$ at 1248 MHz. As such, the fractional circular polarization of the source was less than 10.1\% on 2020 April 11. This upper limit is well below some of the detections reported by \citet{wang2021}, before and after 2020 April 11 (Figure~\ref{fig:fluxvstime}). 

\section{Discussion}
\label{discussion-sec}

\subsection{RM variability}

Compared to the findings of \citet{wang2021}, the data presented in this paper reveal a much larger range of $RM$ variability in \ASKAP, notably changing the sign of $RM$. We also provide the first absolute polarization angle measurement, corrected for Faraday rotation, which may be important to establish whether the source has a stable magnetic axis in the plane of the sky.

The significance of the variable RM is best illustrated by a numeric example. $RM$ is an integral quantity over the complete line of sight, but $RM$ variability with amplitude of order $10^2\ \radm$ is rare. Its origin must be sought in terms of a localized phenomenon associated with the source, as opposed to the ISM on larger scales, since the RM of most sources is not variable.

The distance traveled by relative motion with speed $v$ in one year is $d \approx 10^{-3} v_6$ pc, where $v_6$ has units of 1000 km s$^{-1}$. 
The variation of $RM$ is $\sim 10^2\ \radm\ yr^{-1}$ between 2020 April 11 and 2021 February 9, and $\sim 10^4\ \radm\ yr^{-1}$ between the two measurements by \citet{wang2021} in 2021 February. If this change is effected by relative motion with speed of order $10^3\ \kms$, the line of sight scale of $10^{-3}\ \rm pc$ indicates variation of $n_e B_\|$ in the range $10^5\ \rm cm^{-3}\ \mu G$ to $10^7\ \rm cm^{-3}\ \mu G$.
In the regular warm ionized medium, $n_e \sim 0.1\ \rm cm^{-3}$, $B \sim 5\ \rm \mu G$, a distance $10^{-3}\ \rm pc$ corresponds to a rotation measure increment of order $2 \times 10^{-4}\ \radm$, which is six orders of magnitude smaller than the observed RM variability of \ASKAP. Supernova remnants are believed to have $n_e B_\| \lesssim 10^3$ \citep{2012SSRv..166..231R}, and no supernova remnant is known to be associated with \ASKAP\ (Figure~\ref{fig:ASKAP_pos}). The RM variability on time scales of days to several months suggests that the $RM$ variability arises in an unusual plasma associated with the source.

\subsection{Depolarization by diffractive scattering}

\label{sec:discuss_depol}

\begin{figure*}
    \centering
    \includegraphics[width=\linewidth]{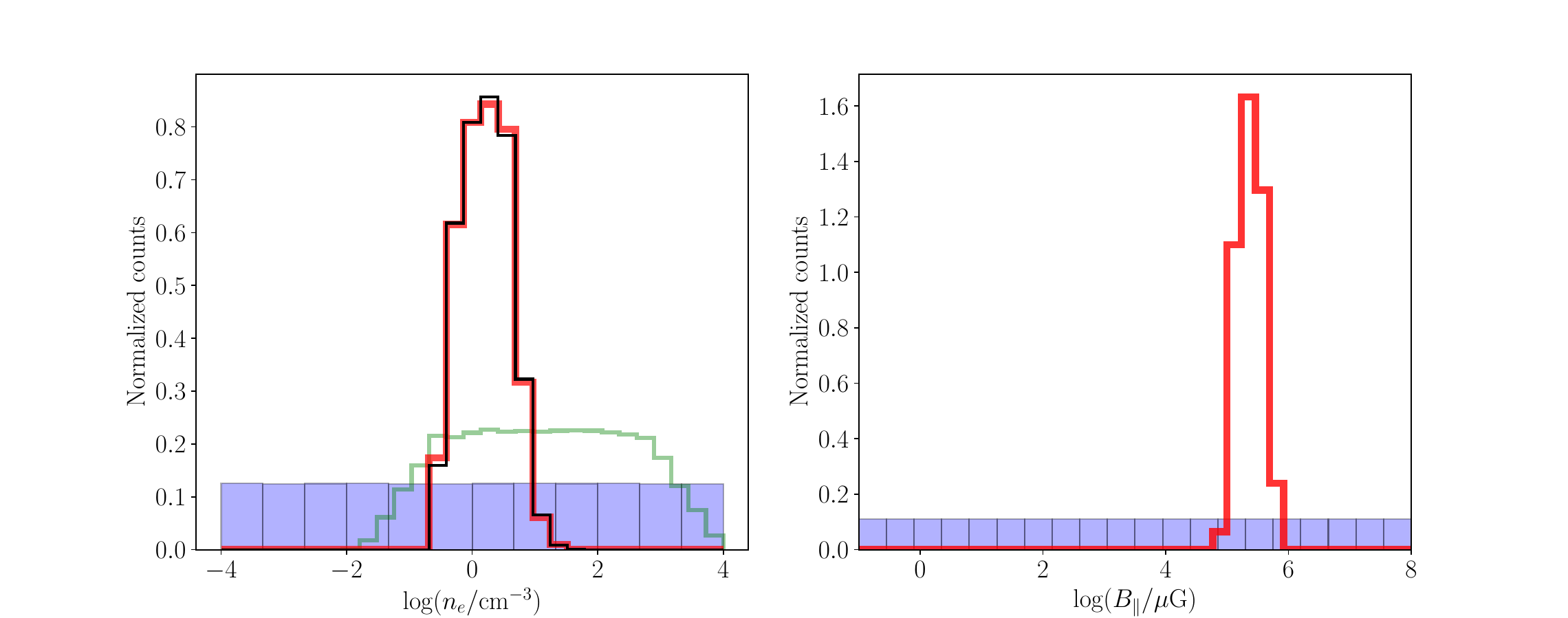}
    \caption{Distributions of $n_e$ (left) and $B_\|$ (right) under the condition that $\nu_{\chi S}$ and $\nu_{rs}$ match the conditions described in the text. The blue histograms show the uniform distributions of the input parameters. The red histograms show the distributions that satisfy the conditions for both $\nu_{\chi S}$ and $\nu_{rs}$. In the left panel, the green histogram shows the distribution that matches only the condition for $\nu_{\chi S}$ and the black histogram shows the distribution that matches only the condition for $\nu_{rs}$. These distributions are not shown on the right, as they would both appear the same as the blue histogram.}
    \label{fig:depol_model}
\end{figure*}

Compared to the $RM$ variability, the data indicate a very modest but non-zero Faraday depth dispersion. Our value is marginally smaller than that of \citet{wang2021} at $\sigma_\phi \lesssim 5.7\ \radm$. Faraday depth dispersion arises from differences in Faraday rotation between lines of sight that are combined within the beam. Instrumental Faraday depth dispersion arising from frequency-averaging the visibilities to 4 MHz channels and fitting Equation~\ref{depol-eq} is less than $0.15\ \radm$ for a source with $RM = 63.9\ \radm$. Equation~\ref{depol-eq} refers, strictly speaking, to a screen of turbulent cells with angular size much smaller than the angular size of the background source. Considering the shortest known variability time scale of \ASKAP, its angular size is of the order of a light day, or 21 milli-arcseconds at the distance of the Galactic center, 8.3 kpc. This is so small that we may suspect that the origin of the observed Faraday depth dispersion is different from the assumptions made in Equation~\ref{depol-eq}.

Diffractive scattering is the distortion of the wave front by an intervening plasma, analogous to the effect of Earth's atmosphere on the light of a star causing seeing and scintillation \citep{1972MNRAS.157...55W, 1990ARA&A..28..561R}. It gives rise to multiple signal paths through the interstellar medium from the source to the observer. Several frequency-dependent effects are observable, such as pulse broadening for a pulsed source and scintillation, which occurs for any source (pulsed or not) with angular size smaller than a minimum angular size defined by the scattering plasma. These multiple signal paths also result in Faraday depth dispersion (c.f. Equation~\ref{RM-eq}). 

Galactic pulsars usually have negligible Faraday depth dispersion \citep{sobey2019}. There are some exceptions, with pulsars displaying wavelength-dependent depolarization comparable to or more than what is measured for \ASKAP\ \citep{sobey2021}. \citet{2009MNRAS.396.1559N} found that the RM of some bright pulsars depends on pulse phase with a range of order $10\ \radm$, and attributed this to scattering. 
The effects of scattering of radio waves on Faraday rotation of a source with very small angular size were recently described by \citet{beniamini2022} in the context of Fast Radio Bursts. We analyze the Faraday depth dispersion of \ASKAP\ in the context of their model.

\citet{beniamini2022} modeled diffractive scattering for FRBs, for which \citet{2022Sci...375.1266F} have presented a relation between pulse broadening and Faraday dispersion. The scattering in these sources occurs in a plasma close to the source, not in the Galactic ISM. Since \ASKAP\ is a compact source in a special environment, we can apply the model of \citet{beniamini2022}. A necessary but insufficient condition for depolarization is that the mean Faraday rotation angle $\Delta \theta \ge 1\ \rm rad$. Taking the RM variability amplitude as a lower limit of the mean RM of the plasma, this condition is satisfied for $\lambda^2 \gtrsim 6.7\times 10^{-3}\ \rm m^2$, or $\nu \lesssim 3.7\ \rm GHz$. \citet{wang2021} found depolarization for $\lambda^2 \gtrsim 0.05\ \rm m^2$, roughly coincident with the emergence of detectable circular polarization at the $\gtrsim 10\%$ level. The total fractional polarization, including Stokes $V$, dropped significantly below 100\%. 

\citet{beniamini2022} defined a critical frequency $\nu_{rs}$ for wavelength-dependent depolarization by a scattering screen, and frequency $\nu_{\chi S}$ below which circular polarization may arise from the scattering. Figure~\ref{fig:depol_model} shows the results of a parameter search that varied the electron density $n_e$, line-of-sight component of the mean magnetic field, $B_\|$, distance $d$ and size of the scattering screen $L$. The distance was varied between 0.1 kpc and 10 kpc, while the size of the scattering screen was varied between $10^{-4}\ \rm pc$ and 1 kpc. The spectral resolution $\mathcal{R}$ was set to the THOR-GC resolution ($4\ \rm MHz$). All other model parameters were kept at the values assumed by \citet{beniamini2022} (their Figure 3). We then constrained the models according to $1.1 < \nu_{\chi S} < 1.3\ \rm GHz$ and $1.5 < \nu_{rs}< 2.5\ \rm GHz$. The distribution of allowed values for $n_e$ and $B_\|$ are shown as red histograms in Figure~\ref{fig:depol_model}. We find that both constraints are met if the density and magnetic field are within a factor $\sim 3$ of $n_e = 2\ \rm cm^{-3}$ and $B_\| = 2 \times 10^5\ \rm \mu G$. Interestingly, the strongest constraint on $n_e$ comes from $\nu_{rs}$, while both constraints are required to place limits on $B_\|$. The depth of the screen and the distance are not constrained by this experiment.

The model of \citet{beniamini2022} thus implies a plasma with density that is normal for the ISM in the Galactic center region \citep{2017ApJ...835...29Y} but a magnetic field that is several orders of magnitude stronger than typical ISM magnetic fields. We find $n_e B_\| \sim 10^5 - 10^6\ \rm cm^{-3} \mu G$, which is roughly consistent with the range estimated from the RM variability. The models approximately but not precisely reproduce the observed polarization of \ASKAP. To give a specific example, assuming $n_e= 1.5\ \rm cm^{-3}$, $B_\| = 3.5 \times 10^5\ \mu$G, $L =  10^{-3}\ \rm pc$, $d = 8.3\ \rm kpc$, velocity of the source $10^3\ \rm km\ s^{-1}$ and eddy velocity $100\ \rm km\ s^{-1}$ yield 1.39 GHz linear polarization 89\%, circular polarization 4.5\%, while at 1.0 GHz, the linear polarization is 76\% and the circular polarization is 17\%. These values approximate the observed fractional polarization and upper limits, but the frequencies $\nu_{\chi S} = 0.91\ \rm GHz$ and $\nu_{rs} = 1.2\ \rm GHz$ appear lower than the observations indicate.

The low-frequency spectral break reported here may be due to synchrotron self-absorption. Following \citet{1981ARA&A..19..373K} for the frequency of the peak brightness of a compact synchrotron source, with $B = 3 \times 10^5\ \rm \mu G$, peak flux density $100\ \rm mJy$, and angular size less than 21 milli-arcseconds, we find the peak brightness occurs at 220 MHz. With the same parameters, but angular size of the source half the upper limit set by causality, the peak would occur near the VLITE frequency. So the parameters we find are consistent with the interpretation of the observed low-frequency spectral break in terms of synchrotron self absorption. It should be kept in mind though, that the emission mechanism in \ASKAP\ may be more complex than the standard assumptions about synchrotron emission.

\subsection{Nature of \ASKAP\ \\ and the scattering screen}

The upper range of magnetic field strength in supernova remnants is believed to be of order $1\ \rm mG$ \citep{2012SSRv..166..231R} and the product $n_e B_\| \sim 10^3\ \rm cm^{-3} \mu G$ is adopted for supernova remnants as environments for some Fast Radio Bursts \citep{2022Sci...375.1266F,2022ApJ...928L..16Y}. The observed linear and circular polarization of \ASKAP\ requires a scattering screen with a magnetic field that is two orders of magnitude stronger, but one that is still weaker by many orders of magnitude than the magnetic field of a typical pulsar $\sim 10^{12}\ \rm G$ \citep[e.g.][]{philippov2022}. Bearing in mind that the emission of \ASKAP\ may not be incoherent synchrotron emission, the synchrotron cooling time for electrons emitting GHz frequency synchrotron emission in a magnetic field of $10^6\ \rm \mu G$ is of the order of a month, which is too long to explain the observed variability on time scales of a day. 

In the context of a compact source behind a scattering screen, a Faraday depth dispersion that is small compared with $RM$ has been suggested as evidence of a large-scale magnetic field, or additional Faraday rotation in a separate plasma along the line of sight. However, when the $RM$ is variable with a large amplitude over a short time scale, we must associate the $RM$ itself with small-scale structure, and its variability with relative motions as outlined above. The implied plasma density $n_e \sim 2\ \rm cm^{-3}$ is also too small to provide significant free-free opacity to explain the low-frequency spectral break reported in Section~\ref{sec:results}. The thermal emission of the scattering screen is undetectable because its filling factor in the synthesized beam is very small.
In summary, we find that the scattering model by \citet{beniamini2022} implies a plasma with unremarkable density but strong magnetic field associated with \ASKAP.

\ASKAP\ may be a high-velocity neutron star moving through the inner Galaxy. The physical scale for structure in the screen, $\sim 10^{-3}\ \rm pc$ is not strongly constrained by the data, but it follows from the RM variability and a plausible but high speed. A highly supersonic neutron star, possibly with a pulsar wind, would create a bow shock structure. The distance between a highly supersonic pulsar and the apex of the contact discontinuity between a pulsar wind and the shocked interstellar medium was given by \citet{2017JPlPh..83e6301K} as
\begin{equation}
R_a = 6.5 \times 10^{16} n^{-1/2} (f_{\Omega} \dot{E}_{36})^{1/2} v_{7}^{-1}\ \rm cm,
\end{equation}
and with $n = 2$ the particle density in $\rm cm^{-3}$, $v_7 = 10$ the speed in $10^7\ \rm cm\ s^{-1}$, $\dot{E}_{36} = 1$ the mechanical luminosity of the pulsar wind and $f_{\Omega} = 1$, a dimensionless anisotropy factor, we find $R_a = 4.6 \times 10^{15}\ \rm cm$ or $R_a = 1.5 \times 10^{-3}\ \rm pc$. The approximate scale of the screen and the relative speed of order $10^{3}\ \rm \kms$ match the shocked interstellar medium in the bow shock of a high-velocity neutron star. The implied high magnetic field strength and modest density could arise in a turbulent wake. 

The size of the radio emission region is less than a light day ($8.4 \times 10^{-4}\ \rm pc$) because of causality constraints and the rapid variability time scale reported by \citet{wang2021}. The emission of \ASKAP\ is more highly polarized than the theoretical maximum for optically thin incoherent synchrotron emission, which has fractional polarization $\lesssim 70\%$ to $\lesssim 85\%$ \citep[higher if the power law electron energy spectrum is steeper][]{1977OISNP..89.....P}. Our data show that the source reached a maximum brightness in 2020 before dimming by a factor $\sim 30$ and brightening again to the 2021 peak observed by \citet{wang2021}. This supports the idea of occasional brief enhanced injection (acceleration) of relativistic electrons into the emission region and that the variable, steep to ultra-steep, spectral index is related to the time of an observation since the injection of relativistic electrons.  

The origin of the bursts requires speculation on the reason for brief intermittent injection (acceleration) of relativistic electrons. The time scale between peak brightness appears to be months to years, suggested by a combination of modest sampling in time, detectable low-level emission in the period between the 2020 and 2021 peaks, and the fact that \ASKAP\ was not observed before 2020. \citet{wang2021} did not detect pulsed emission but their observations were inconclusive because of the variability of the source. Absence of pulsation of (most of) the burst emission is significant because the physical scales derived from the variability allow an emission region that is orders of magnitude larger than the radius of the light cylinder, $R_{\rm LC} = 1.5 \times 10^{-9} P\ \rm pc$ with $P$ the pulsar's rotation period in seconds.

The variability of \ASKAP\ is unlike normal pulsars, so the mechanism of the bursts may be related to special conditions. It is interesting that the upper limit to the size of the emission region is comparable to the distance between the neutron star and the vertex point of the contact discontinuity between the pulsar wind and the swept-up medium. This leaves the possibility of injection of new particles into the magnetosphere because of a change in environment, for example if \ASKAP\ runs into a structure in the interstellar medium that changes the density or relative velocity of the surrounding medium. For a time scale of 1 year between observed bursts, and a relative velocity of $10^3\ \rm \kms$, the scale of a plasma structure would be of order $1.0 \times 10^{-3}\ \rm pc$ or $\sim 200\ \rm AU$. Such plasma structures are known to exist from extreme scattering events \citep[e.g.][]{2015ApJ...808..113C}. The physical scale is also similar to the size of the Solar heliopause. 

If \ASKAP\ is approximately at the distance of the Galactic center, it must be located well within the Fermi and eROSITA bubbles \citep{2020ApJ...894..117Z}. The environment of \ASKAP\ may be very different from the interstellar medium in the solar neighborhood, with a higher density of gas and stars, possibly stirred up by activity of Sgr A* within the past $\lesssim 3$ Myr \citep{2012ApJ...756..181G,2022NatAs...6..584Y}. This makes it plausible that the environment of \ASKAP\ changes on time scales shown in Figure~\ref{fig:fluxvstime} and required by the $RM$ variability, when moving at its implied high speed.

The above picture of a highly supersonic pulsar running into a plasma structure, or perhaps the outskirts of another solar system, suggests it may be a rare kind of source with activity on a time scale of a year that may come to an end. Magnetohydrodynamic simulations of supersonic pulsars \citep{2019MNRAS.484.4760B,2020JPhCS1623a2002B} model the interaction region as a steady, anisotropic pulsar wind that is generated at the light cylinder. Emission from the much larger interaction region between the pulsar wind and the interstellar medium is not expected to be pulsed. We note that much older pulsars with a negligible pulsar wind may be better described by a magnetosphere interacting with the interstellar medium, in analogy to a planetary magnetosphere that interacts with the solar wind.

Direct observations of rapid evolution of the spectral index and possible time evolution of the spectral break reported here would provide valuable information on the emission mechanism and the particle acceleration process. This can be achieved by continued monitoring of this source and a more complete statistical sample of Galactic radio transients. The latter is a prospect for new radio telescopes with high survey speed, such as the Square Kilometre Array.

\section{Summary and Conclusions}

\ASKAP\ was detected with the VLA on 2020 April 11 with flux density $20.6 \pm 1.1\ \rm mJy$ at 1.23 GHz and spectral index $\alpha = -3.1 \pm 0.2$ (THOR-GC) and 339 MHz flux density $122.6 \pm 20.4$ mJy (VLITE). The linear polarization at 1.23 GHz was $76.7 \pm 3.9\%$ and a $3\sigma$ upper limit for the circular polarization $|V|/I < 10.1\%$ was found.

On 2020 April 11 the rotation measure of \ASKAP\  was $63.9 \pm 0.3\ \radm$. The rotation measure of \ASKAP\ is more variable than previously thought, and it changes sign. A basic geometric argument of relative motion of a plasma and a compact source suggests that the Faraday rotation occurs in a plasma with $n_{e} B_{\|}$ of order $10^5$ to $10^7\ \rm cm^{-3}\ \mu G$, which is several orders of magnitude higher than the warm ionized ISM (WIM) and also 2 orders of magnitude beyond the range of $n_e B_\|$ thought to be representative for supernova remnants.

The simultaneous THOR-GC L-band and VLITE 339 MHz data reveal a low-frequency break in the spectrum of \ASKAP. If this break is caused by free-free absorption, the free-free opacity at 1 GHz is $\tau = 0.4 \pm 0.06$. The thermal emission of such a plasma can only remain undetected if the filling factor within our beam is much smaller than 1. Such a plasma would almost certainly be associated with the source. However, free-free absorption is not consistent with our analysis of a scattering screen with modest density $n_e \sim 2\ \rm cm^{-3}$. The spectral break may be the result of synchrotron self-absorption in a source with angular size less than 21 milli-arcseconds, magnetic field $\sim 3 \times 10^5\ \rm \mu G$ and peak flux density $100\ \rm mJy$ at a few hundred MHz.

The in-band depolarization of \ASKAP\ is quantified by Faraday depth dispersion $\sigma_\phi = 4.8^{+0.5}_{-0.7}\ \radm$, marginally smaller than the value reported by \citet{wang2021}. If this Faraday dispersion arises from scattering of radio waves in a plasma, the model by \citet{beniamini2022} suggests the scattering plasma has a density that is comparable to WIM density near the Galactic center, but strongly magnetized with magnetic field of the order of $3 \times 10^5\ \rm \mu G$.

We conclude that the variable $RM$ and the Faraday depth dispersion of \ASKAP\ are consistent with the presence of a highly magnetized plasma associated with the source. This kind of plasma may be found in the wake of a high-velocity neutron star interacting with its environment.

\section*{acknowledgments}
The National Radio Astronomy Observatory is a facility of the National Science Foundation operated under cooperative agreement by Associated Universities, Inc. Construction and installation of VLITE was supported by the Naval Research Laboratory Sustainment Restoration and Maintenance fund. $RM$-Tools is maintained by the Canadian Initiative for Radio Astronomy Data Analysis (CIRADA; cirada.org). This work is based on data from eROSITA, the soft X-ray instrument aboard SRG, a joint Russian-German science mission supported by the Russian Space Agency (Roskosmos), in the interests of the Russian Academy of Sciences represented by its Space Research Institute (IKI), and the Deutsches Zentrum für Luft- und Raumfahrt (DLR). The SRG spacecraft was built by Lavochkin Association (NPOL) and its subcontractors, and is operated by NPOL with support from the Max Planck Institute for Extraterrestrial Physics (MPE). The development and construction of the eROSITA X-ray instrument was led by MPE, with contributions from the Dr. Karl Remeis Observatory Bamberg \& ECAP (FAU Erlangen-Nuernberg), the University of Hamburg Observatory, the Leibniz Institute for Astrophysics Potsdam (AIP), and the Institute for Astronomy and Astrophysics of the University of Tübingen, with the support of DLR and the Max Planck Society. The Argelander Institute for Astronomy of the University of Bonn and the Ludwig Maximilians Universität Munich also participated in the science preparation for eROSITA. J.M.S. acknowledges the support of the Natural Sciences and Engineering Research Council of Canada (NSERC), 2019-04848, and a generous private donor of a substantial server for data processing. M.R.R is a Jansky Fellow of the National Radio Astronomy Observatory, USA. W.M.P., K.E.N., and T.E.C. acknowledge that basic research in radio astronomy at the U.S.\ Naval Research Laboratory is supported by 6.1 Base funding. This work was performed in part at the Jet Propulsion Laboratory, California Institute of Technology, under contract with the National Aeronautics and Space Administration (80NM0018D0004). S.A.D. acknowledge the M2FINDERS project from the European Research Council (ERC) under the European Union's Horizon 2020 research and innovation programme (grant No 101018682). M.C.S acknowledges financial support from the Royal Society (URF\textbackslash R1\textbackslash 221118) and the European Research Council under the ERC Starting Grant ``GalFlow" (grant No 101116226).

\bibliography{Bibliography}{}
\bibliographystyle{aasjournal}

\end{document}